\documentclass[a4paper]{article}  
\usepackage{indentfirst}  
\usepackage{bm}    
\usepackage{graphicx}  
\usepackage{graphicx}
\usepackage{enumitem}
\usepackage{subfigure}
\usepackage{amsmath}
\usepackage{amssymb}
\usepackage{multirow}
\usepackage[table,xcdraw]{xcolor}
\usepackage{url}
\usepackage{fancyhdr}

\usepackage{geometry}
\begin{document}    
\thispagestyle{empty} \vspace*{0.8cm}\hbox
to\textwidth{\vbox{\hfill\noindent \\ \textit{Proceedings of the 8th International Conference on Pedestrian and Evacuation Dynamics (PED2016)\\
Hefei, China - Oct 17 -- 21, 2016\\
Paper No. 84}
\hfill}}
\par\noindent\rule[3mm]{\textwidth}{0.2pt}\hspace*{-\textwidth}\noindent
\rule[2.5mm]{\textwidth}{0.2pt}

\begin{center}
\LARGE\bf Avoid or Follow? Modelling Route Choice Based on Experimental Empirical Evidences
\end{center}

\begin{center}
\rm Luca Crociani$^{1}$\footnote{Corresponding author.} \ Daichi Yanagisawa$^2$ \ Giuseppe Vizzari$^1$ \\ Katsuhiro Nishinari$^2$ and Stefania Bandini$^{1,2}$
\end{center}

\begin{center}
\begin{small} \sl
${}^{\rm 1}$ Complex Systems and Artificial Intelligence research center, University of Milano--Bicocca \\Viale Sarca 336 -- Ed. U14, 20126 Milano (ITALY) \\   
\{name.surname\}@disco.unimib.it \\
${}^{\rm 2}$ Research Center for Advanced Science and Technology, The University of Tokyo \\4-6-1 Komaba, Meguro-ku, Tokyo 153-8904 (JAPAN) \\ 
tDaichi@mail.ecc.u-tokyo.ac.jp; tknishi@mail.ecc.u-tokyo.ac.jp\\
\end{small}
\end{center}
\vspace*{2mm}

\begin{center}
\begin{minipage}{15.5cm}
\parindent 20pt\small
\noindent\textbf{Abstract -} Computer-based simulation of pedestrian dynamics reached meaningful results in the last decade, thanks to empirical evidences and acquired knowledge fitting fundamental diagram constraints and space utilization. Moreover, computational models for pedestrian wayfinding often neglect extensive empirical evidences supporting the calibration and validation phase of simulations. The paper presents the results of a set of controlled experiments (with human volunteers) designed and performed to understand pedestrian's route choice. The setting offers alternative paths to final destinations, at different crowding conditions. Results show that the length of paths and level of congestion influence decisions (negative feedback), as well as imitative behaviour of ``emergent leaders'' choosing a new path (positive feedback). A novel here illustrated model for the simulation of pedestrian route choice captures such evidences, encompassing both the tendency to avoid congestion and to follow emerging leaders. The found conflicting tendencies are modelled with the introduction of a utility function allowing a consistent calibration over the achieved results. A demonstration of the simulated dynamics on a larger scenario will be also illustrated in the paper.
\end{minipage}
\end{center}

\begin{center}
\begin{minipage}{15.5cm}
\begin{minipage}[t]{2.3cm}{\bf Keywords:}\end{minipage}
\begin{minipage}[t]{13.1cm}
Pedestrian Dynamics, Route Choice, Experiment, Simulation, Validation
\end{minipage}\par\vglue8pt
\end{minipage}
\end{center}

\section{Introduction}  
The simulation of pedestrians and crowds is a consolidated research and application area that, although existing results are currently employed on a daily basis by designers and decision makers, still presents challenges for researchers in different fields and disciplines. Innovations in the field come from diverse contributions, from attempts to integrate  automated analysis of crowd behaviour and simulation~\cite{DBLP:journals/expert/VizzariB13} to investigations on the impact of emotional aspects on crowd dynamics~\cite{DBLP:journals/aamas/BosseHKTWW13}, and even on the usage of simulation as a real-time support to crowd management~\cite{DBLP:journals/sj/TsiftsisGS16}.

Even if we only consider choices and actions related to walking in normal situations, modelling human decision making activities and actions is a complicated task. Decisions on pedestrian's actions are taken at different levels of abstraction, from path planning to the regulation of distance from other pedestrians (potentially having different kinds of relationships, as recent results on the impact of groups indicate~\cite{10.1371/journal.pone.0121227}) and obstacles present in the environment. A factor making the evaluation of simulation results more complicated is the fact that the measure of success and validity of a model is not the \emph{optimality} of pedestrian dynamics with respect to some cost function (either on the individual or on the aggregated dynamics), as (for instance) in robotics. The success of a simulation model is associated to its \emph{plausibility}, the adherence of the simulation results to data that can be acquired by means of observations or experiments.

The present research effort is aimed at producing insights on this aspect: an experiment involving pedestrians has been set up to investigate to which extent pedestrians facing a relatively simple choice (i.e. choose one of two available gateways leading to the same target area) in which, however, they can face a trade-off situation between length of the trajectory to be covered and estimated travel time. The closest gateway, in fact, is initially selected by most pedestrians but it is too narrow to allow a smooth passage of so many pedestrians, and it quickly becomes congested. The other choice can therefore become much more reasonable, allowing a higher average walking speed and comparable (if not even lower) travel time. We observed that several pedestrians choose longer paths to preserve high walking speed, and often do so following a first \emph{emerging leader}. Modelling this kind of choices with current approaches can be problematic.

The model presented in this paper represents a step in the direction fitting this kind of evidences but providing a general approach to this kind of pedestrian decisions. The proposed model encompasses, in fact, both a proxemic tendency to avoid congestion, as well as an \emph{imitation} mechanism: these conflicting tendencies can be calibrated according to empirical evidences. The former aspect represents a sort of negative feedback caused by crowding conditions (i.e. avoid a congested exit) but, on the other hand, the incentive to following another pedestrian that just decided to choose a more distant but less crowded passage represents a positive feedback on this choice. In the following, we will first describe the experiment and the achieved results. The description of the model will be given in Section~3, followed by the discussion of its calibration and achieved results. 

\section{Experimental Study}\label{sec:exp}
\begin{figure}
\begin{center}
\includegraphics[width=.6\columnwidth]{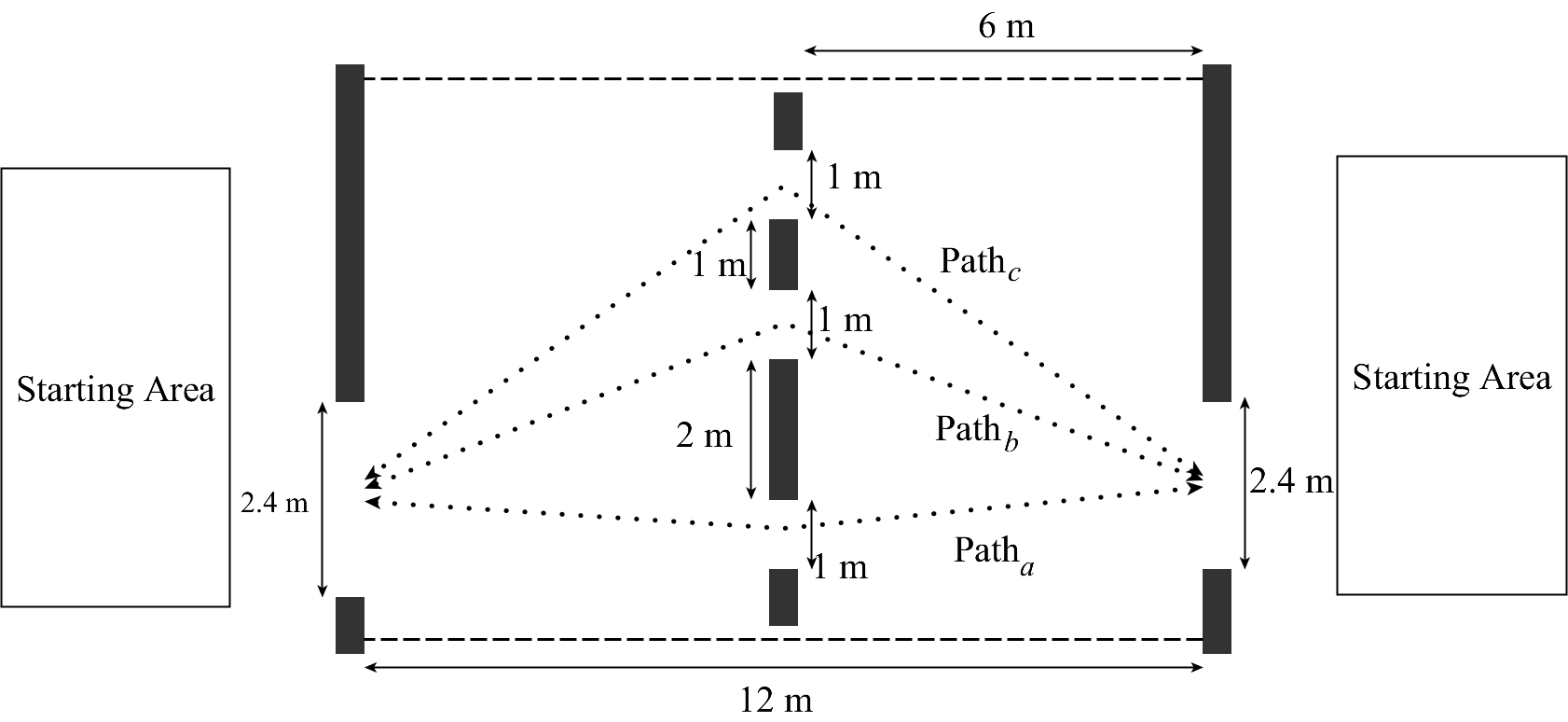}
\caption{Configuration of the setting for the experiment.}\label{fig:setting}
\end{center}
\end{figure}

The experiment has been performed at the University of Tokyo in November 2015. A group of 46 persons has participated, uniformly composed of male students aged around 20 years old. The setting has been configured with the intention to acquire empirical evidences on the influence of crowding conditions on route choice decisions. The setting is designed to describe an elementary choice: it is characterized by a rectangular environment of $7.2\times12$ m$^2$ divided in two areas of equal size; the access is regulated by three \emph{gates} positioned to create three paths of different lengths. A schematic representation of the scenario is illustrated in Figure~\ref{fig:setting}.

The entrance and exits of the environment are aligned on the x-axis, in order to generate a shortest path (path$_{a}$) and to induce the decisions of the participants. The average lengths of the three possible paths --calculated as the sum of distances between the central points of the crossed openings-- is as following: (i) path$_{a}$ 12.08; (ii) path$_{b}$ 12.85; (iii) path$_{c}$ 14.76. 

The difference between Path$_{a}$ and Path$_{b}$ is relatively small ($<$ 1 m) while Path$_{c}$ is significantly longer than the shortest path. These differences will be reflected on the achieved results pointing out, as generally known, that the main element influencing the route choice is the distance. 
After each run, the participants were asked to regroup in the starting area right after the exit point, in order to optimize the procedural times. This has also allowed to balance the potential influence of the cultural elements related to left--right preference of movement, by reversing the direction of flow in the scenario. 
The openings related to Path$_{b}$ and Path$_{c}$ were eventually closed to configure different procedures. Each procedure has been repeated 4 times to achieve a more consistent dataset. The procedures are described in the following:

\begin{enumerate}
\item only Path$_{a}$ allowed;
\item Path$_{a}$ and Path$_{b}$ allowed;
\item Path$_{a}$ and Path$_{c}$ allowed;
\item all gates are open.
\end{enumerate}

The procedure iterations have been performed with a randomized schedule, to possibly avoid bias provided by the learning of participants. Finally, to stimulate the will to minimize the travelling time towards the destination, the participants have been asked to reach \emph{quickly} the opposite side of the setting. 

\subsection{Methodology for the Analysis}
The experiment has been performed under an arcade of the university buildings, due to weather conditions. The lack of safe points to attach any camera to the ceiling did not allow to have a zenithal perspective in the video recordings, which would be useful for an automatic tracking. Hence, four HD cameras positioned on tall tripods (5m circa) have been used to record the experiment. The videos have been synchronized by means of a global chronometer shown to the cameras at the beginning of the recording. Since the frame rate of the two cameras with the highest resolution was slightly variable (a small number of frames was skipped during the recording), some manual adjustment has been done to the video synchronization. This was possible by using visible events in the overlapping areas of multiple camera views and also by using the audio track of the videos. The software \emph{AviSynth} has been employed to achieve the synchronization and merge the multiple video tracks. The results that will be presented in the following subsection has been achieved by means of manual counting and tracking.

\subsection{Results and Discussion}

The video footages of the performed experiments did not allow us to perform a fine tracking of the exact trajectories followed by the different pedestrians, mainly due to the positioning of cameras. Nonetheless, different types of analysis will be carried out to acquire novel empirical evidences on pedestrian route choice decisions in presence of crowding conditions. 

So far, we focused on a relatively simple analysis that regarded the number of participants who passed through each gate depending on the experimental procedure, and therefore on the level of congestion. Our aim was to verify if the experimental results supported the conjecture that pedestrians, when the perceived level of congestion makes less appealing the shortest path, choose a longer trajectory to preserve their walking speed. Additional analyses will be carried out, within the limits posed by the vantage point, as a consequence of the results of this first round of integrated analysis and synthesis activities~\cite{DBLP:journals/expert/VizzariB13}.

The Table~\ref{tab:countingGate} shows the counting of people passing through each gate for procedures with multiple open gates. As already briefly discussed, procedure 2 and 3 are characterized by a decision among two choices of different lengths; in procedure 2 the difference among the alternatives is quite small (i.e. less than a meter, less than $10\%$ of the shortest trajectory, considering the middle point of the gates for the measurement), whereas in procedure 3 the longer trajectory is more significantly worse than the shortest path (i.e. over $20\%$ longer). Finally, procedure 4 allows pedestrians to choose among all three alternatives.

Intermediate gates are relatively small (1 m) but the evidence shown that they still allow the passage of two pedestrians almost at the same time. Moreover, the passage between the starting area and the region before the gates is relatively wide (2.4 m) and here pedestrians are able to walk side by side; this implies that some of them are naturally closer to the best trajectory while for others the worst ones are not that longer.

Results are essentially in line with the expectations on the conjecture that pedestrians would distribute among the alternative passages in case of congestion. In fact, in procedure 2, almost half of the pedestrians chooses the slightly longest path to avoid the congested path. The fact that, in procedure 3, the longer alternative ($Path_c$) is more significantly worse, makes this choice appealing to a lower number of pedestrians. Procedure 4, finally, shows that the $Path_a$ and $Path_b$ are perceived as almost equivalent, but a few pedestrians even choose $Path_c$ in order to preserve a high walking speed by avoiding congestion on the best alternatives.

\begin{figure*}[t]
\begin{center}
\subfigure[19 s]{\includegraphics[width=.3\columnwidth]{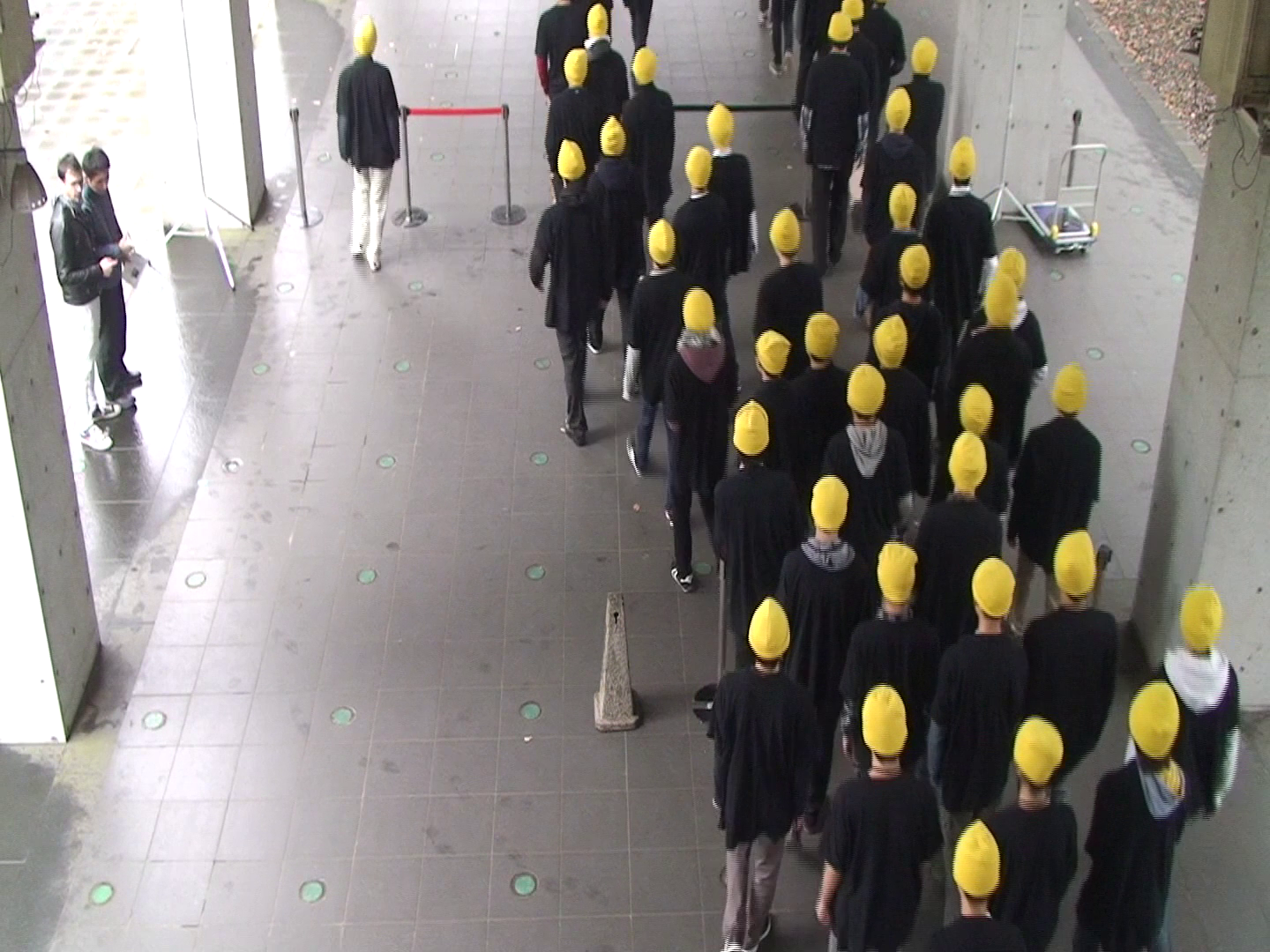}}\hspace{.8cm}
\subfigure[21 s]{\includegraphics[width=.3\columnwidth]{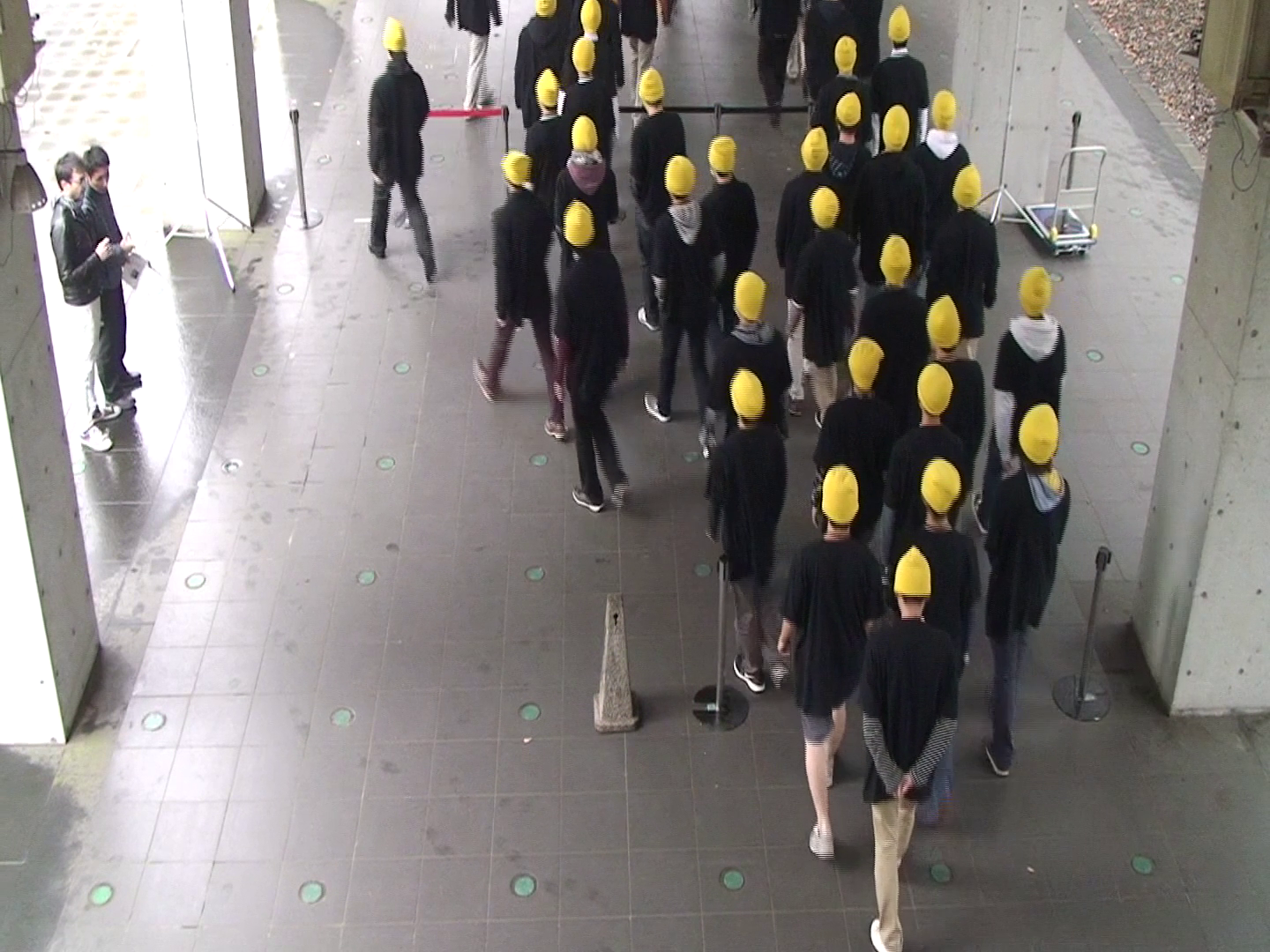}}\\
\subfigure[23 s]{\includegraphics[width=.3\columnwidth]{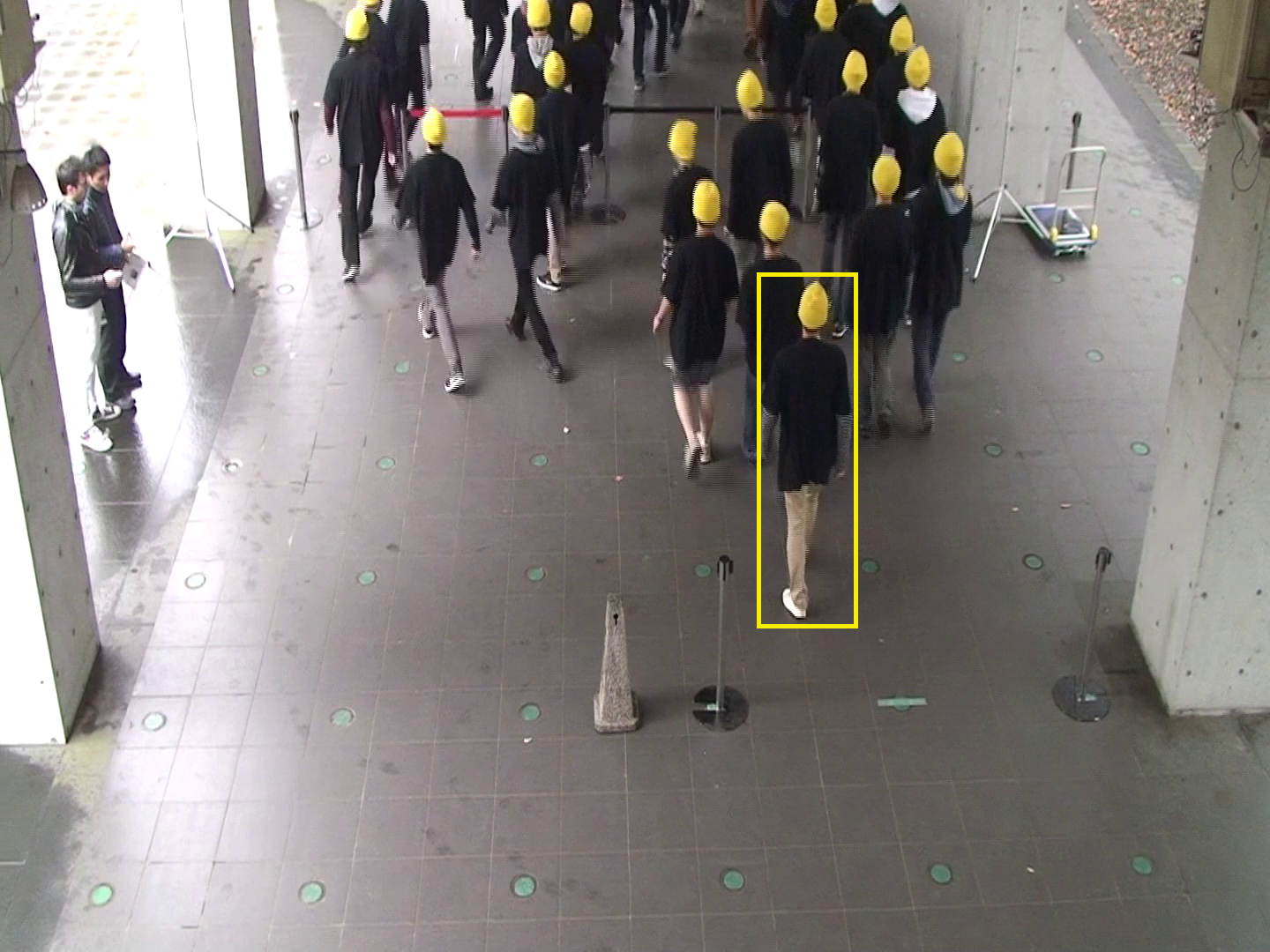}}\hspace{.8cm}
\subfigure[25 s]{\includegraphics[width=.3\columnwidth]{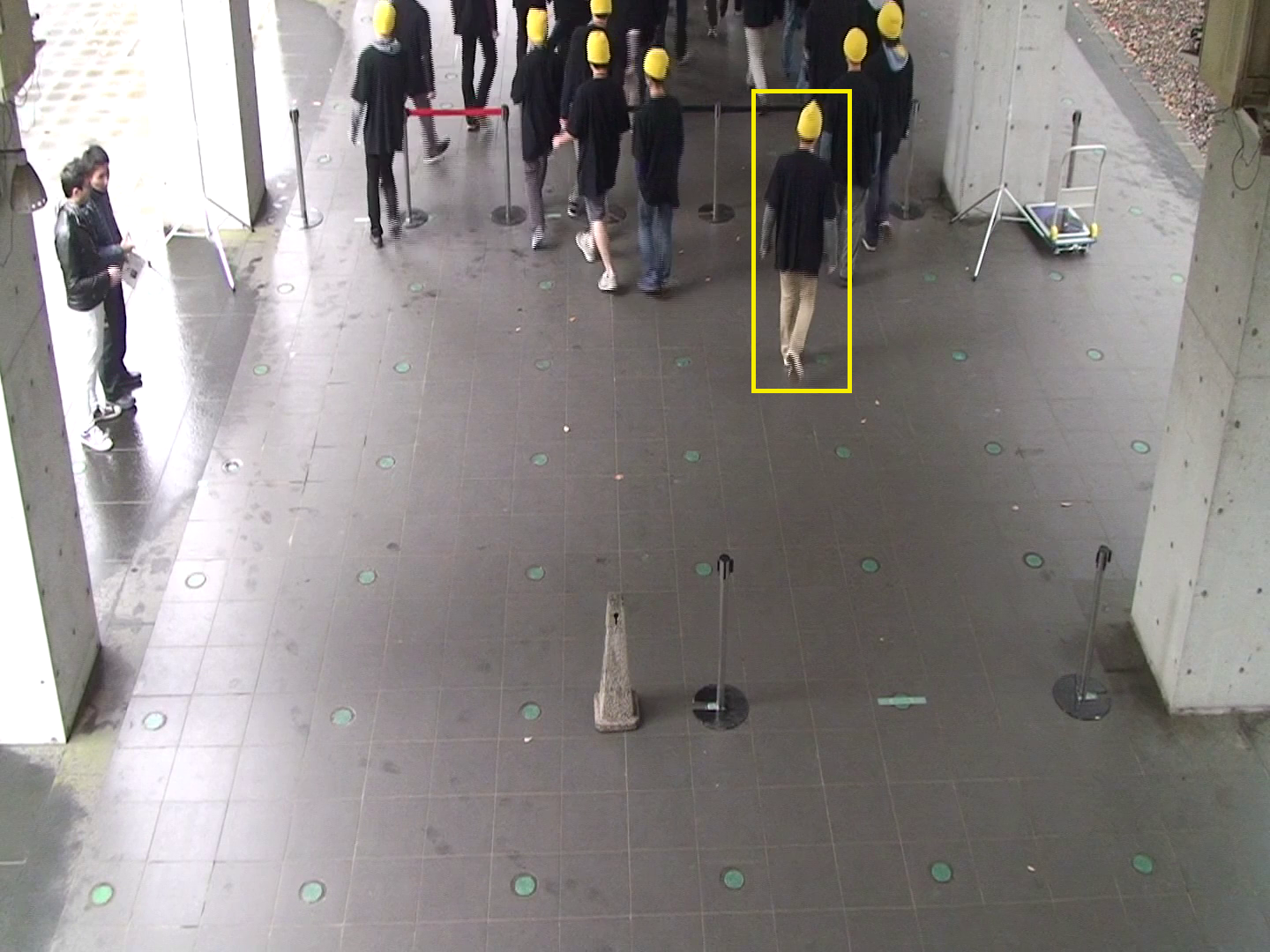}}
\caption{Video camptures from one camera during procedure 4. The behaviours considered in this paper can be recognized: at 19 s of the video, the crowding of $Path_a$ and $Path_b$ leads a person to employ $Path_c$. After 2 s, other students have taken the same decision, successively followed by others. In the bottom images, a decision change is apparent: the highlighted person is firstly headed to the central gate, but after a short while he notices that $Path_a$ is getting empty again and he employs it.}\label{fig:exp_screenshots}
\end{center}
\end{figure*}

In particular, a few general qualitative considerations on all the analyzed procedures can be done:

\begin{itemize}

\item pedestrians choosing longer paths ($Path_b$ and $Path_c$) generally enter the area before the gates on upper side, referring to the map of the scenario in Figure~\ref{fig:setting};

\item pedestrians choosing longer paths generally do so after some preceding ones have perceivably chosen the best path, and therefore can be considered as potential future competitors either for the occupation of the gate or for the path leading to it; this is particularly apparent for procedure 3 and 4;

\item sometimes, when choosing the longest path ($Path_c$) pedestrians seem to \emph{follow} someone before them that had chosen this trajectory before and, at the same time, \emph{avoiding} much closer pedestrians that have perceivably chosen other shorter paths. 

\end{itemize}

Figure~\ref{fig:exp_screenshots} emphasizes these considerations by showing 4 sequential screen-shots from one iteration of procedure 4.  While these results are still quite aggregated, they suggest that pedestrians are driven by the expected travel time rather than just the length of the trajectory. Moreover, the observed following dynamics could be explained as a form of cooperative behaviour by which those who are on the back 
select a preceding pedestrian whose choice is assumed to be promising, at a glance, without actually computing the expected quickest time to the exit. To some extent, this reminds conflicting but simultaneously present behavioural components of \emph{cohesion} and \emph{separation}, considered in simulation models~\cite{Moussaid2010}. In particular, the modelling approach that will be introduced in the following Section, will consider both the fact that other pedestrians are generally perceived as repulsive but also the fact that the decision of a pedestrian to detour (i.e. change a previous decision on the path to be followed) is a \emph{locally perceivable event} that might trigger a similar reconsideration by nearby pedestrians.

\begin{table}[t]
\begin{center}
\scriptsize
\begin{tabular}[c]{||c|c|c|c||}
\hline
\textbf{Procedure} & \textbf{Path$_a$} & \textbf{Path$_b$} & \textbf{Path$_c$} \\
\hline
\hline
2 & 22 & 24 & 0\\
& 23 & 23 & 0\\
& 25 & 21 & 0\\
& 23 & 23 & 0\\
\hline
Average & 23.25 & 22.75 & 0\\
\hline
3 & 27 & 0 & 19\\
& 28 & 0 & 18\\
& 30 & 0 & 16\\
& 27 & 0 & 19\\
\hline
Average & 28 & 0 & 18\\
\hline
4 & 18 & 16 & 12\\
& 22 & 19 & 5\\
& 21 & 18 & 7\\
& 22 & 19 & 5\\
\hline
Average & 20.75 & 18 & 7.25\\
\hline
\end{tabular}
\caption{Counting of number of people passed through each path for procedures 2, 3 and 4.}\label{tab:countingGate}
\end{center}
\end{table}

\section{A Novel Model for Pedestrian Wayfinding}\label{sec:model}
The proposed agent-based model defines two components of the agent, respectively dedicated to the low level reproduction of the movement towards a target (i.e. the operational level, considering a three level model described in~\cite{Hoogendoorn2004}) and to the decision making activities related to the next destination to be pursued (i.e. the wayfinding). Due to lack of space, the mechanisms for the physical reproduction of movement will not be described\footnote{For technical details of this aspect it is referred to~\cite{CrocianiJCA}.} and the discussion will be limited to the different representations of the environment composing the knowledge needed for the wayfinding. 

\subsection{The Representation of the Environment and the Knowledge of Agents}
The environment~\cite{WeynsDefJaamas} is discrete and modelled with a rectangular grid of 40 cm sided square cells, as usually applied in this context. The simulation scenario is drawn by means of several \emph{markers}. Basic markers of the scenario are \emph{start areas}, \emph{obstacles} and \emph{final destinations} --ultimate targets of pedestrians. To allow the wayfinding, two other markers are introduced. \emph{Openings} are sets of cells that divide the environment into regions, together with obstacles. These objects constitutes decision elements for the route choice, which will be denoted as \emph{intermediate targets}. Finally, \emph{regions} are markers that describe the type of the region: with them it is possible to design particular classes of regions (e.g. stairs, ramps).

This model uses the \emph{floor fields} approach~\cite{Burstedde2001}, spreading potentials from cells of obstacles and destinations to provide information about distances. The two types of floor fields are denoted as \emph{path field}, spread from each target, and \emph{obstacle field}, a unique field spread from all obstacle cells. In addition, a \emph{dynamic} floor field that has been denoted as \emph{proxemic field} is used to reproduce plausible distance towards the agents in low density situations.

The presence of intermediate targets allows the computation of a graph-like representation of the walkable space, based on the concept of \emph{cognitive map}~\cite{Tolman1948}. The algorithm for the computation of this data structure is defined in~\cite{crociani2014hybrid} and it uses the information of the floor fields associated to openings and final destinations. Recent approaches explores also the modelling of partial and incremental knowledge of the environment by agents (e.g.~\cite{Andresen2016}), but this aspect goes beyond the scope of the current work. The data structure identifies \emph{regions} (e.g. a room) as nodes of the labelled graph and \emph{openings} as edges. Overall the cognitive map allows the agents to identify their position in the environment and it constitutes a basis for the generation of an additional knowledge base, which will enable the reasoning for the route calculation. 

This additional data structure is denoted as \emph{Paths Tree} and it contains the free flow travel times\footnote{Calculated with their \emph{desired speed} of walking.} related to \emph{plausible} paths towards a final destination, from each intermediate target in the environment. The concept of plausibility of a path is encoded in the algorithm for the computation of the tree, which is thoroughly discussed in~\cite{DBLP:journals/ia/CrocianiPVB15}. 

Formally, for the choice of paths the agents access the information of a Paths Tree, generated from a final destination $End$, with the function $Paths(R, End)$. Given the region $R$ of the agent, this outputs a set of couples $\{(P_i,tt_i)\}$. $P_i = \left\lbrace \Omega_k, \ldots, End \right\rbrace$ is the ordered set describing paths starting from $\Omega_k$, belonging to $Openings(R)$ and leading to $End$. $tt_i$ is the respective free flow travel time.

\subsection{The Route Choice Model of Agents}

This novel part of the model is inspired by the behaviour observed in the experiment. The aim is to propose an approach that would enable agents to choose their path considering distances as well as the evolution of the dynamics, being able to follow emerging leaders as well as to avoid congestions. At the same time, the model must provide a sufficient variability of the results and a calibration over possible empirical data.

\begin{figure}[t]
\begin{center}
\includegraphics[width=.7\columnwidth]{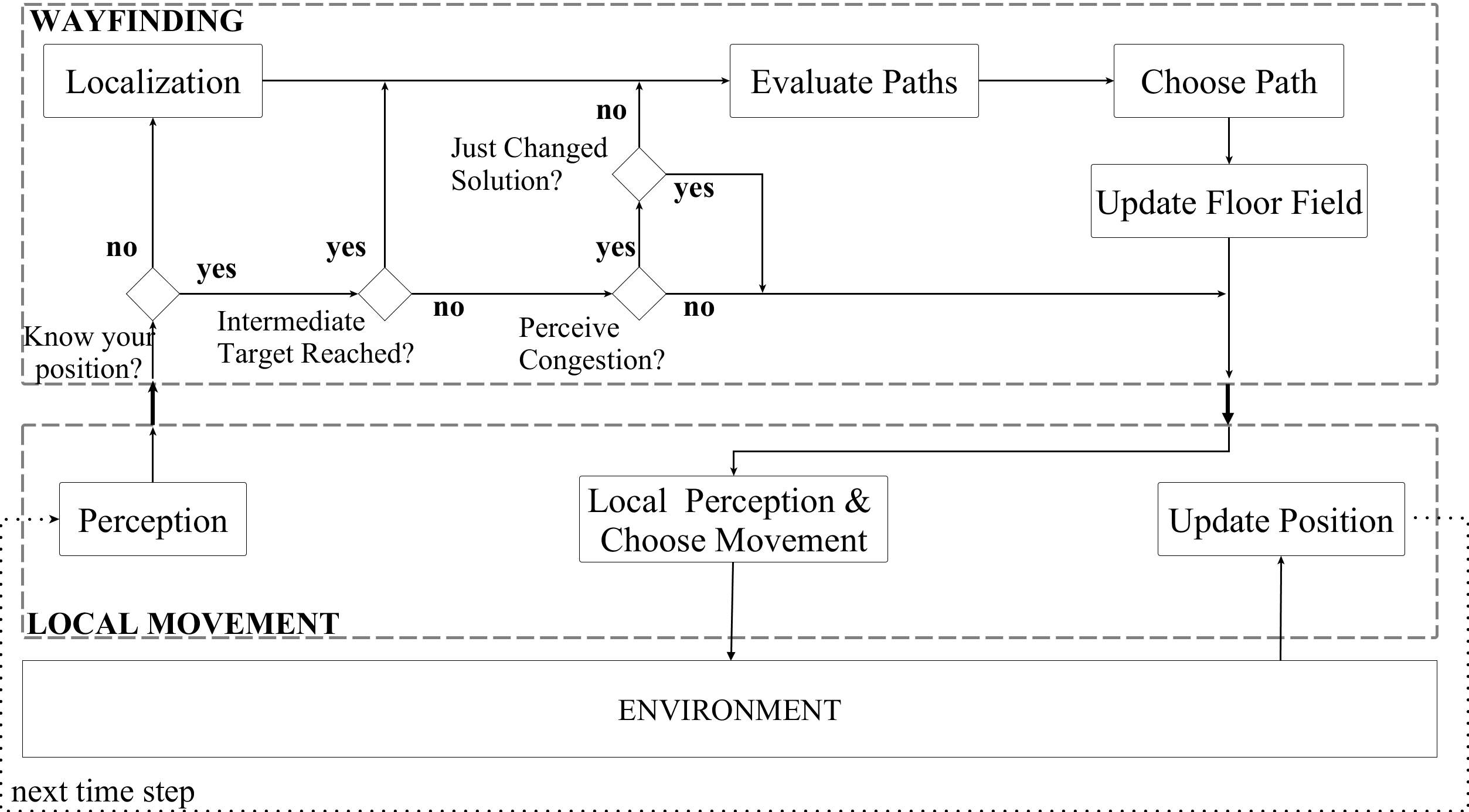}
\caption{The life-cycle of the agent, emphasizing the two components of the model.}\label{fig:agent_lifeCycle}
\end{center}
\end{figure}

The life-cycle of the agent is illustrated in Figure~\ref{fig:agent_lifeCycle}. First of all, the agent performs a perception of his situation, aimed at understanding its position in the environment and the perceivable markers from its region. At the very beginning of its life, the agent does not have any information about its location, thus the first assignment is its \emph{localization}, inferring the location in the Cognitive Map. Once the agent knows its region, it loads the Paths Tree and evaluates the possible paths towards its final destination. 

In this model, the evaluation and choice of paths is designed with the concept of \emph{path utility}, computed to assign a probability to be chosen. The probabilistic choice outputs a new intermediate target of the agent, which will be followed at the operational level. The functions introduced for the wayfinding model are \emph{Evaluate Paths} and \emph{Choose Paths}, which will be now discussed.

\subsubsection{The Utility and Choice of Paths}
The function that computes the probability of choosing a path is exponential with respect to the utility value associated to it. This is analogous to the choice of movement at the operational layer:

\begin{equation}
Prob(P) = N \cdot \mbox{exp}\left(\kappa_{tt} Eval_{tt}(P) - \kappa_q Eval_{q}(P) + \kappa_f Eval_{f}(P)\right)
\end{equation}

The exponential function is used to emphasize differences in the perceived utility values of paths, limiting the choice of relatively bad solutions. The exponent comprises the three observed components influencing the route choice decision, which are aggregated with a weighted sum. In particular, the first element evaluates the expected travel times, the second considers the \emph{queuing} conditions through the considered path and the last one introduces a positive influence of choices of nearby agents to pursue the associated path (i.e. imitation of emerging leaders). Each of the three functions provide normalized values within the range $[0,1]$. 

\subsubsection{The Evaluation of Travelling Times}
Distances and travel times are elements that significantly affect our choice of route and this is observed also in the experiment, where the shortest path resulted as the most chosen on average. First of all, the free flow travel time $tt_i$ of a path $P_i$ is achieved with $Paths(R, End)$ ($End$ is the agent's final destination, identifying the appropriate Paths Tree, and $R$ is the region of the agent) and it is summed with the free flow travel time to reach the first opening $\Omega_k$ described by each path:

\begin{equation}
\mathit{TravelTime}(P_i) = \mathit{tt}_i + \frac{PF_{\Omega_k}(x,y)}{\mathit{Speed}_d}
\end{equation}

$PF_{\Omega_k}(x,y)$ is the value of the path field associated to $\Omega_k$ in the position of the agent and $\mathit{Speed}_d$ is its \emph{desired speed}. The value of the travelling time is then evaluated by means of the following function:
 
\begin{small}
\begin{equation}
\mathit{Eval}_{tt}(P)= N_{tt} \cdot \frac{\min\limits_{\scriptscriptstyle P_i \in \mathit{Paths}(r)} (\mathit{TravelTime}(P_i))}{\mathit{TravelTime}(P)}
\label{eq:Eval_tt_2}
\end{equation}
\end{small}

$N_{tt}$ is the normalization factor. The minimum value of travel times provides a range of the function of (0,1], being 1 for the path with minimum travel time and decreasing as the difference with the other paths increases. This modelling choice, makes this function describe the \emph{utility} of the route in terms of travel times, instead of its \emph{cost}.

\subsubsection{The Evaluation of Congestion}
The competition arisen between the participants of the experiment made some of them choose longer paths, in order to avoid the possible queue. The behaviour modelled in the agent reflects this effect, by considering congestion as a negative element for the evaluation of the path. For the evaluation of this component, a function is firstly introduced for denoting agents $a'$ that precedes the evaluating agent $a$ in the route towards the opening $\Omega$ of a path $P$:

\begin{small}
\begin{equation}
\mathit{Forward}(\Omega ,a) =  |\{a' \in Ag\backslash \{a\} : Dest(a') = \Omega\ \land \mathit{PF}_{\Omega}(\mathit{Pos}(a')) < \mathit{PF}_{\Omega}(\mathit{Pos}(a))\}|
\end{equation}
\end{small}

where $Pos$ and $Dest$ indicates respectively the position and current destination of the agent; the fact that $\mathit{PF}_{\Omega}(\mathit{Pos}(a')) < \mathit{PF}_{\Omega}(\mathit{Pos}(a))$ assures that $a'$ is closer to $\Omega$ than $a$,. Each agent is therefore able to perceive the main direction of the others (its current destination). This kind of perception is plausible considering that only preceding agents are counted, but we want to restrict its application when agents are sufficiently close to the next passage. To introduce a way to calibrate this perception, the following function and an additional parameter $\gamma$ is introduced:

\begin{small}
\begin{equation}
\mathit{PerceiveForward}(\Omega ,a) =
\begin{cases}
\mathit{Forward}(\Omega, a), & \mbox{if } \mathit{PF}_{\Omega}(\mathit{Pos}(a))< \gamma \\
0, & \mbox{otherwise}
\end{cases}
\end{equation}
\end{small}

The function $\mathit{Eval}_q$ is finally defined with the normalization of $\mathit{PerceiveForward}$ values for all the openings connecting the region of the agent:

\begin{small}
\begin{equation}
\mathit{Eval}_q(P) = N_q \cdot \frac{\mathit{PerceiveForward}(\mathit{FirstEl}(P), \mathit{myself})}{\mathit{width}(\mathit{FirstEl}(P))}
\end{equation}
\end{small}

where $\mathit{FirstEl}$ returns the first opening of a path, $\mathit{myself}$ denotes the evaluating agent and $\mathit{width}$ scales the evaluation over the width of the door. 

\subsubsection{Propagation of Choices - Following behaviour}
Another behaviour observed in the experiment described a substantial following of emerging leaders who sudden change their route due to the surrounding crowding conditions.  The final component of this model aims at representing the effect of additional stimulus for the agents, generated by decision changes of other persons. An additional grid is then introduced to model this event, whose functioning is similar to the one of a dynamic floor field. The grid, called \emph{ChoiceField}, is used to spread a gradient from the positions of agents that, at a given time-step, change their plan due to the perception of congestion.

The functioning of this field is described by two parameters $\rho_c$ and $\tau_c$, which defines the diffusion radius and the time needed by the values to \emph{decay}. The diffusion of values from an agent $a$, choosing a new target $\Omega'$, is performed in the cells $c$ of the grid with $\mathit{Dist}(\mathit{Pos}(a),c) \leq \rho_c$ with the following function:

\begin{small}
\begin{equation}
\mathit{Diffuse}(c,a) =
\begin{cases}
1/\mathit{Dist}(\mathit{Pos}(a),c) & \mbox{if } \mathit{Pos}(a) \neq c\\
1 & \mbox{otherwise}
\end{cases}
\end{equation}
\end{small}

The diffused values persist in the \emph{ChoiceField} grid for $\tau_c$ simulation steps, then they are simply discarded. The index of the target $\Omega'$ is stored together with the diffusion values, thus the grid contains in each cell a vector of couples $\{(\Omega_m, \mathit{diff}_{\Omega_m}), \ldots, (\Omega_n, \mathit{diff}_{\Omega_n})\}$, describing values of influence associated to each opening of the related region. While multiple neighbour agents change their choices towards the opening $\Omega'$, the values of the diffusion are summed up in the respective $\mathit{diff}_{\Omega'}$. Moreover, after having changed its decision, an agent spreads the gradient in the grid for a configurable amount of time steps represented by an additional parameter $\tau_{a}$, thus it influences the choices of its neighbours for $\tau_{a}$ time.

The existence of values $\mathit{diff}_{\Omega_k} > 0$ for some opening $\Omega_k$ implies that the agent is influenced in the evaluation phase by one of these openings, but the probability for which this influence is effective is, after all, regulated by the utility weight $\kappa_f$. In case of having multiple $\mathit{diff}_{\Omega_k} > 0$ in the same cell, a individual influence is chosen with a simple probability function based on the normalized weights $\mathit{diff}$ associated to the cell. Hence, for an evaluation performed by an agent $a$ at time-step $t$, the utility component $Eval_{f}$ can be equal to 1 only for one path $\overline{P}$, between the paths having $\mathit{diff}_{\Omega_k} >0$ in the position of $a$.

\section{Calibration of the Model and Conclusions}
\begin{figure}[t!]
\begin{center}
\subfigure[Procedure 2]{\includegraphics[width=.55\textwidth]{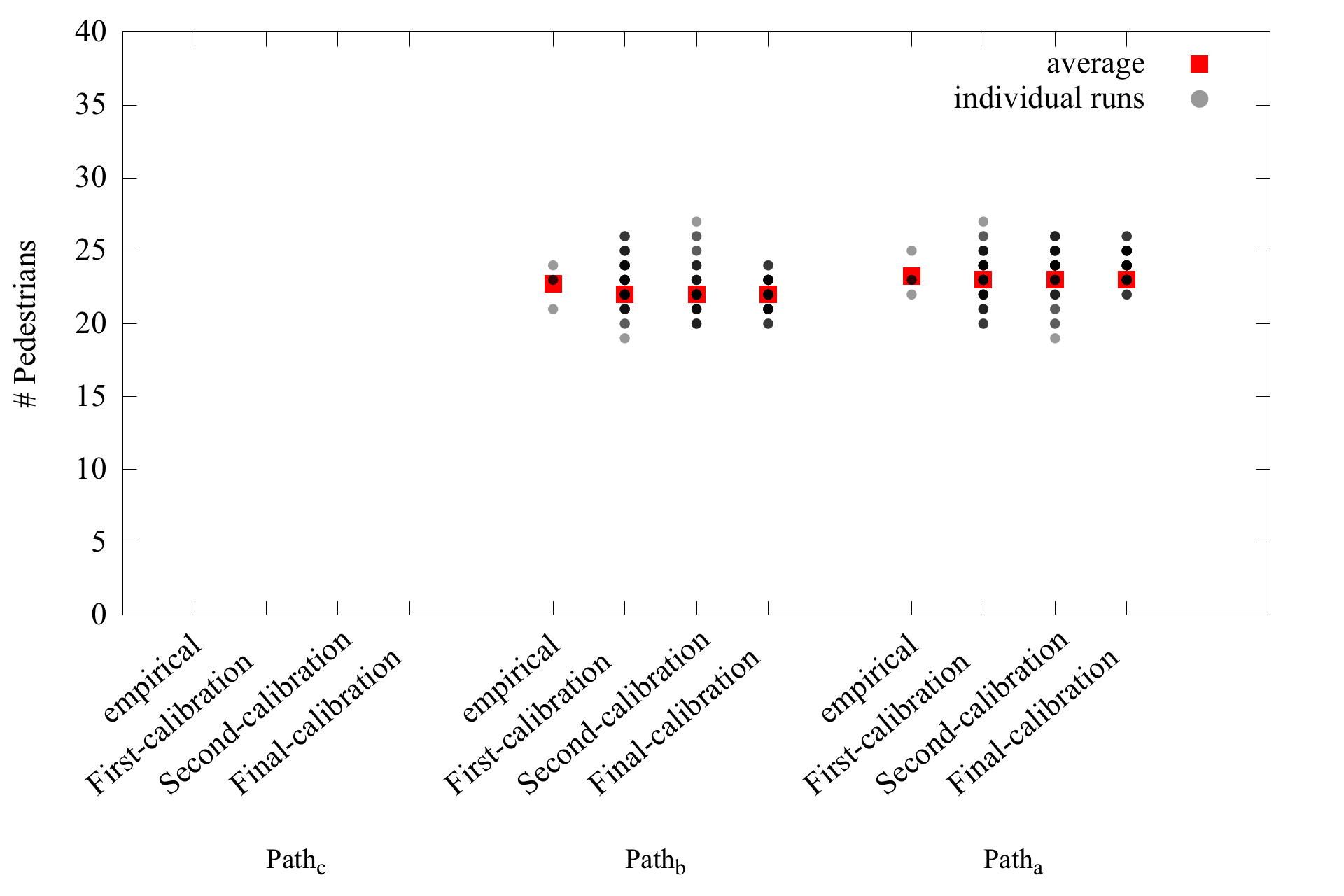}\label{fig:Tokyo_exp2}}
\subfigure[Procedure 3]{\includegraphics[width=.55\textwidth]{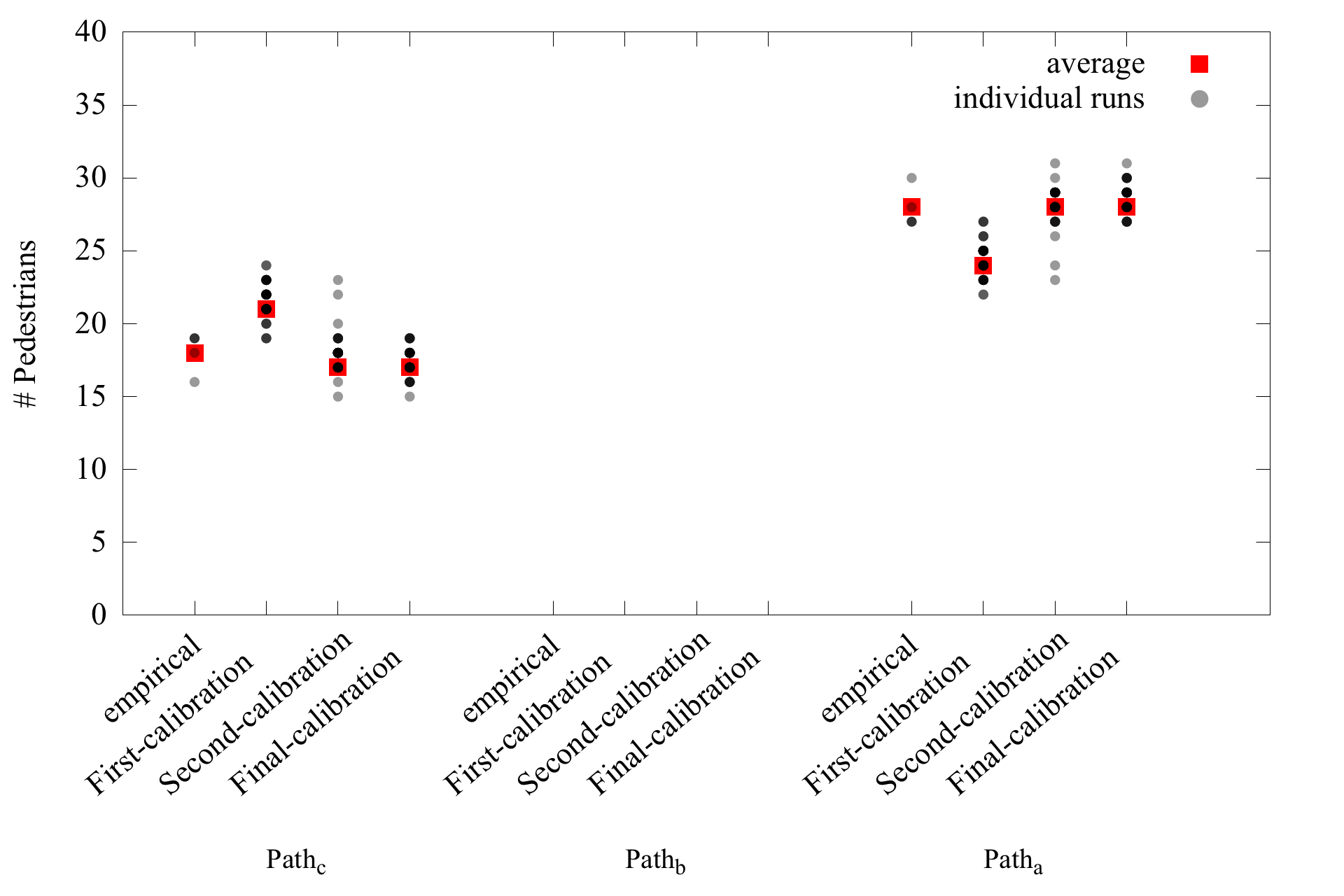}\label{fig:Tokyo_exp3}}
\subfigure[Procedure 4]{\includegraphics[width=.55\textwidth]{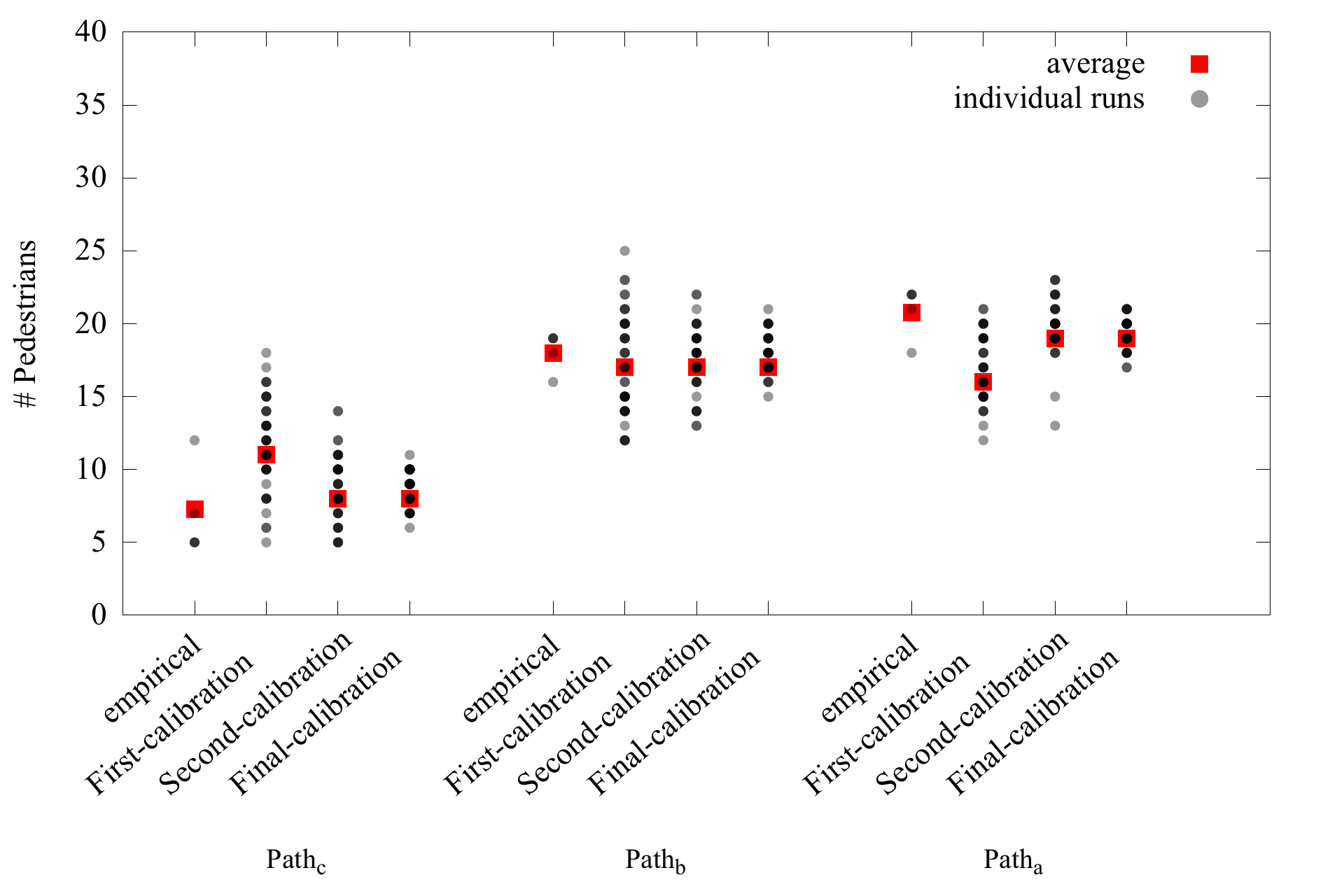}\label{fig:Tokyo_exp4}}
\caption{Comparison between empirical data and simulation results with different calibrations.}\label{fig:simResults}
\end{center}
\end{figure}

To evaluate the reliability of the proposed model, we employed the observed data of the experiment for performing a comparison with results coming from the simulation of the same setting. A set of 50 simulation iterations per each experimental procedure has been configured and the counting of agents passing through each gate has been gathered. Results are shown in Figure~\ref{fig:simResults}.

The value for parameters has two kinds of impact on the achieved results: it obviously determines the number of pedestrians choosing a specific door, but it also affects the variance of these numbers. In fact, moving from the first calibration ($\kappa_{tt} = 10, \kappa_{q}=7, \kappa_{f}=5$) to the second ($\kappa_{tt} = 10, \kappa_{q}=2.5, \kappa_{f}=0.5$), we achieved average results closer to the empirical observation, however the variability was too high. This lead us to considering a final calibration ($\kappa_{tt} = 100, \kappa_{q}=25, \kappa_{f}=5$) that preserved the proportion among the weights, but increasing their values to reduce the effect of randomness, leading therefore to very similar average results but a much lower variability, more similar to the one observed in the experiments.

In conclusion, the presented model is effective in reproducing the observed phenomena, but it is of general applicability and we are in the process of evaluating the effects of its usage in larger scale egress scenarios.

\subsubsection*{Acknowledgement} 
The experiment was authorised and supported by The University of Tokyo. The authors would like to thank Francesco Crippa for his support in the development and experimentation of the model. 



\begin{thebibliography}{10}
\providecommand{\url}[1]{#1}
\csname url@samestyle\endcsname
\providecommand{\newblock}{\relax}
\providecommand{\bibinfo}[2]{#2}
\providecommand{\BIBentrySTDinterwordspacing}{\spaceskip=0pt\relax}
\providecommand{\BIBentryALTinterwordstretchfactor}{4}
\providecommand{\BIBentryALTinterwordspacing}{\spaceskip=\fontdimen2\font plus
\BIBentryALTinterwordstretchfactor\fontdimen3\font minus
  \fontdimen4\font\relax}
\providecommand{\BIBforeignlanguage}[2]{{%
\expandafter\ifx\csname l@#1\endcsname\relax
\typeout{** WARNING: IEEEtran.bst: No hyphenation pattern has been}%
\typeout{** loaded for the language `#1'. Using the pattern for}%
\typeout{** the default language instead.}%
\else
\language=\csname l@#1\endcsname
\fi
#2}}
\providecommand{\BIBdecl}{\relax}
\BIBdecl

\bibitem{DBLP:journals/expert/VizzariB13}
G.~Vizzari and S.~Bandini, ``Studying pedestrian and crowd dynamics through
  integrated analysis and synthesis,'' \emph{{IEEE} Intelligent Systems},
  vol.~28, no.~5, pp. 56--60, 2013.

\bibitem{DBLP:journals/aamas/BosseHKTWW13}
\BIBentryALTinterwordspacing
T.~Bosse, M.~Hoogendoorn, M.~C.~A. Klein, J.~Treur, C.~N. van~der Wal, and
  A.~van Wissen, ``Modelling collective decision making in groups and crowds:
  Integrating social contagion and interacting emotions, beliefs and
  intentions,'' \emph{Autonomous Agents and Multi-Agent Systems}, vol.~27,
  no.~1, pp. 52--84, 2013. [Online]. Available:
  \url{http://dx.doi.org/10.1007/s10458-012-9201-1}
\BIBentrySTDinterwordspacing

\bibitem{DBLP:journals/sj/TsiftsisGS16}
\BIBentryALTinterwordspacing
A.~Tsiftsis, I.~G. Georgoudas, and G.~C. Sirakoulis, ``Real data evaluation of
  a crowd supervising system for stadium evacuation and its hardware
  implementation,'' \emph{{IEEE} Systems Journal}, vol.~10, no.~2, pp.
  649--660, 2016. [Online]. Available:
  \url{http://dx.doi.org/10.1109/JSYST.2014.2370455}
\BIBentrySTDinterwordspacing

\bibitem{10.1371/journal.pone.0121227}
\BIBentryALTinterwordspacing
N.~W.~F. Bode, S.~Holl, W.~Mehner, and A.~Seyfried, ``Disentangling the impact
  of social groups on response times and movement dynamics in evacuations,''
  \emph{PLoS ONE}, vol.~10, no.~3, pp. 1--14, 03 2015. [Online]. Available:
  \url{http://dx.doi.org/10.1371%2Fjournal.pone.0121227}
\BIBentrySTDinterwordspacing

\bibitem{Moussaid2010}
M.~Moussa{\"{\i}}d, N.~Perozo, S.~Garnier, D.~Helbing, and G.~Theraulaz, ``The
  walking behaviour of pedestrian social groups and its impact on crowd
  dynamics.'' \emph{PloS one}, vol.~5, no.~4, p. e10047, 2010.

\bibitem{Hoogendoorn2004}
S.~P. Hoogendoorn and P.~H.~L. Bovy, ``{Pedestrian route-choice and activity
  scheduling theory and models},'' \emph{Transportation Research Part B:
  Methodological}, vol.~38, no.~2, pp. 169--190, 2004.

\bibitem{CrocianiJCA}
S.~Bandini, L.~Crociani, and G.~Vizzari, ``An approach for managing
  heterogeneous speed profiles in cellular automata pedestrian models,''
  \emph{Journal of Cellular Automata}, (in press).

\bibitem{WeynsDefJaamas}
D.~Weyns, A.~Omicini, and J.~Odell, \emph{Autonomous Agents Multi-Agent
  Systems}, vol.~14, no.~1, pp. 5--30, 2007.

\bibitem{Burstedde2001}
C.~Burstedde, K.~Klauck, A.~Schadschneider, and J.~Zittartz, ``{Simulation of
  pedestrian dynamics using a two-dimensional cellular automaton},''
  \emph{Physica A: Statistical Mechanics and its Applications}, vol. 295, no. 3
  - 4, pp. 507--525, 2001.

\bibitem{Tolman1948}
E.~C. Tolman, ``{Cognitive maps in rats and men.}'' \emph{Psychological
  review}, vol.~55, no.~4, pp. 189--208, 1948.

\bibitem{crociani2014hybrid}
L.~Crociani, A.~Invernizzi, and G.~Vizzari, ``{A hybrid agent architecture for
  enabling tactical level decisions in floor field approaches},''
  \emph{Transportation Research Procedia}, vol.~2, pp. 618--623, 2014.

\bibitem{Andresen2016}
E.~Andresen, D.~Haensel, M.~Chraibi, and A.~Seyfried, ``Wayfinding and
  cognitive maps for pedestrian models,'' in \emph{Proceedings of Traffic and
  Granular Flow 2015 (TGF2015)}.\hskip 1em plus 0.5em minus 0.4em\relax
  Springer, (in press).

\bibitem{DBLP:journals/ia/CrocianiPVB15}
L.~Crociani, A.~Piazzoni, G.~Vizzari, and S.~Bandini, ``When reactive agents
  are not enough: Tactical level decisions in pedestrian simulation,''
  \emph{Intelligenza Artificiale}, vol.~9, no.~2, pp. 163--177, 2015.

\end{thebibliography}
\end{document}